# Software Defects Prediction Using Machine Learning

Mitt Shah, Nandit Pujara

*Computer Science and Engineering*

*Nirma University*

Ahmedabad, India

18bce121@nirmauni.ac.in, 18bce130@nirmauni.ac.in

*Abstract*—Software quality is one of the essential aspects of a software. With increasing demand, software designs are becoming more complex, increasing the probability of software defects. Testers improve the quality of software by fixing defects. Hence the analysis of defects significantly improves software quality. The complexity of software also results in a higher number of defects, and thus manual detection can become a very time-consuming process. This gave researchers incentives to develop techniques for automatic software defects detection. In this paper, we try to analyze the state of the art machine learning algorithms' performance for software defect classification. We used seven datasets from the NASA promise dataset repository for this research work. The performance of Neural Networks and Gradient Boosting classifier dominated other algorithms.

## I. INTRODUCTION

With growing demand and technology, the software industry is rapidly evolving. Since humans do most of the software development, defects will inevitably occur. In general, defects can be defined as undesired or unacceptable deviations in software documents, programs, and data [1].Defects may exist in requirements analysis because the product manager misinterprets the customer's needs, and as a result, this defect will also carry on to the system design phase. Defects may also occur in the code due to inexperienced coders. Defects significantly impact software quality, such as increased software maintenance costs, especially in healthcare, and aerospace software defects can have serious consequences. If the fault is detected after deployment, it causes an overhead on the development team as they need to re-design some software modules, which increases the development costs. Defects are nightmares for reputed organizations. Their reputation is affected due to customer dissatisfaction and hence reduces its market share.

Therefore, software testing has become one of the main focuses of industrial research [2]. With the rise in software development and software complexity, the number of defects has increased to the extent that traditional manual methods consume much time and become inefficient. The rise of machine learning has made automatic classification of defects a research hotspot. In this paper, we initially discuss software defects in detail and their different categories available in the literature and then discuss the manual classification methods proposed by various researchers. Finally, we present the analysis of the state of the art machine learning algorithms for automatic software detection.

## II. CONCEPTS AND OVERVIEW OF SOFTWARE DEFECTS

### A. Concept of software defects

Since analysts often can't distinguish between software defects and programming faults, errors, and failure, this article utilizes IEEE 729-1983 (Standard Glossary of Software Engineering Terminology) to characterize defects as, From the inside of the product, the defects are mistakes and errors in the maintenance or development of the product item. From an external perspective, a defect is the violation or failure of the framework/system to accomplish specific capacities [3, 4]. The description of the concepts that are easily mistaken with defects is as follows

1. *Fault:* The software doesn't perform according to the client's expectations and runs in an unsuitable internal state. We can view it as a defect that can prompt software errors and is regarded as dynamic behavior.
2. *Failure:* It refers to the outputs that the software generates at runtime, which the client doesn't accept. For instance, if the

execution capacity is lost, and the client's capabilities are not met, the framework can't meet the fixed asset's execution necessities.
3. *Error:* It is introduced by individuals and changed over into faults under specific conditions. It exists in the whole software life cycle, including error information in the software design, data structure, code, requirements analysis, and other carriers [5].

The quality of software relies upon the number of defects. An excessive number of defects lead to reduced client satisfaction, consuming organization assets and expenses, and slower testing. To spare the costs, improving test productivity is critical to managing defects.

*B. Main research directions of software defects*
1. *Software defect management*

Defect management mainly refers to the collection, statistics, and useful recording of defects. To improve management productivity, engineers have designed many robotized defects management devices. At present, the industry's commonly used tools are mainly JIRA dispatched by Atlassian and Bugzilla, an open-source bug tracking framework provided by Mozilla. These two tools record the transactions, attributes, statistical information of defects, and don't have a more profound investigation and explicit grouping of defects. Defect analysis and classification are significant segments of defect management. Therefore, examination and arrangement of deformities require further exploration of the data recorded in JIRA and Bugzilla.

2. *Software defect analysis*

Software designers regularly use defect analysis to better access programming quality and development quality. Software defect analysis is a strategy for characterizing imperfections and mining the reasons for defects. The motivation behind software defect analysis is to enable analysts to find, locate, evaluate, and improve test efficiency. The defects analysis methods are mostly partitioned into qualitative analysis, quantitative analysis, and attribute analysis [4]. Qualitative analysis strategies mostly incorporate Root Cause Analysis (RCA) and Software Fault Tree Analysis (SFTA). Attribute analysis is commonly partitioned into single attribute analysis and multi-attribute analysis.

3. *Software defect classification*

Software defects are different and complex. A more clear grouping and conglomeration of deficiencies can assist programmers with assessing programming quality, improve analyzers' work productivity, and decrease the trouble of analysis. Classification is likewise useful to suggest repair techniques and reuse test cases [2]. It can comprehend the distribution of defects as per the classification and analysis results, prevents frequently occurring software defects, extraordinarily improve the software development cycle, and in this way, improve the quality of software [6, 7]. In this way, software defects classification is a significant piece of software defect analysis. The outcomes of defect classification directly influences the defect analysis process, so defect classification has great significance. Up until now, software defect classification can be partitioned into manual classification and programmed/automatic classification.

1. *Manual classification of software defects*: Software defect manual classification implies that examiners utilize their insight to group defects into various classes. To start with, the researchers set the ideal classification of defects. They then discover the fault and type match the defect based on experience. Nonetheless, the classification cycle of this strategy is quite complicated and requires a large team. Because of restricted human energy and memory, a lot of data analysis will bring about a lower classification speed, far lower than the computer, and consequently burn-through a great deal of time and assets.
2. *Automatic classification of software defects:* To reduce development costs and improve development productivity, individuals are more inclined to use computers to automatically classify defects. Specialists are attempting to locate a straightforward method to classify defects, and the ascent of AI and machine learning has made the automatic classification of defects a hotspot for industrial research.

III. SOFTWARE DEFECTS MANUAL CLASSIFICATION METHOD

Defect classification is the basis for the quantitative analysis of defects. Due to the numerous components that cause defects, every individual distinguishes the defect type differently, for example, database type, code type, operation type, etc. Each type can keep on being classified; for example, code types can be partitioned into task errors, variable definition errors, etc. There is a wide range of classification methods. Different classification methods should be utilized for various analysis purposes, so the classification methods are also quite complicated. A few common manual classification methods are presented below:

### A. Orthogonal Defect Classification

T. J. Watson of IBM presented another classification technique in 1992: Orthogonal Defect Classification (ODC) [8]. IBM accepts that defect classes ought to be orthogonal to one another, there should be no convergences, and every classification is autonomous of one another. There should not exist a defect that can have a place with one classification that can likewise have a place with another classification. Besides, classes should cover all defects in a comprehensive manner, and there should be no defects that can't be allocated to any of the categories. Since the primary reason for ODC is to improve the development process, this strategy classifies defects in detail, covers all phases of software development, and applies to any software product ODC utilizes numerous attributes to describe defective qualities. Its most recent version defines eight defect attributes [8]. As indicated by these traits' qualities, software defects are divided into the following eight classes: assignment, verification, algorithm, timing, interface, function, association, and documentation

### B. Standard classification for Anomalies

The IEEE standard classification for anomalies [9] was proposed by the IEEE to give a uniform classification standard to software anomalies. These standard traverses the whole software life cycle and effectively distinguishes and tracks defects and improves the development cycle.

The standard defines Failure classification and Defect classification. Before classifying, individuals must decide if to use the Failure classification strategy or the Defect classification technique initially, select various attributes for classification, and break down the classification results. Anomaly classification classifies exceptions into eight classifications: computational problems, interface/timing problems, logic problems, data problems, data processing problems, document quality problems, documentation problems, and reinforcement problems. Each class can likewise be further divided into more detailed subclasses.

This technique is scalable and relies upon the projects' necessities. Because of this, the strategy is more subject to the experience and level of the executive. However, individual subjective factors will seriously influence the classification impact.

### C. Thayer classification

Thayer et al. classify errors as per their inclination by using error reports filled by researchers during testing and clients' feedback [10, 11]. This technique divides errors into logic errors, calculation errors, data processing errors, I/O errors, support software errors, operating systems, interface errors, configuration errors, preset database errors, user demand changes, repeated errors and, global variable errors, requirement implementation error, document error, personnel operation error, and unknown property error. They can likewise be partitioned into 164 subcategories.

This strategy analyzes software errors, yet additionally includes system errors and personnel operation errors. It is cautiously classified and widely used. It is useful for developers to alter the errors. In any case, this technique doesn't consider the reasons for the error and, thus, is unacceptable for improving the software process.

### D. Roger classification

According to the causes of defects, Roger's classification divides defects into 12 categories: Intentional deviation specifications (IDS), incomplete or erroneous specifications (IES) [12]. Misunderstandings communication with customers (MCC), and so on. This classification strategy is generally straightforward, the defect information is shallow, the standard is rough, and the analyzability isn't high, so the technique analyses the reasons behind the method/strategy's introduction.

### E. Defect prevention classification of IBM

In 1990, R. G. Mays et al. at the IBM Research Center proposed the IBM defect prevention classification scheme. This technique is fairly like the STFA, which utilizes causal analysis to classify defects and recursively until they can't be classified once more, considering the defect introduction time. This strategy classifies defect categories into education, negligence, text errors, improper communication, etc. This strategy is likewise vulnerable to abstract factors that lead to various analysis results for various individuals.

### F. Putnam classification

Since the defects exist in the entire cycle of software development, it isn't comprehensive to consider just the errors in the coding stage, and the error attributes produced in each phase of the requirement analysis and system design are different, so it is inaccurate to classify all phases with coding errors classification models. Because of this. Putnam et al. [13] analyzed numerous defects that set forward the classification technique for Putnam's defects. Focusing on the various attributes of various defects in the development cycle, according to the introduction time of defects, the defects are partitioned into six arrangements: system design defects, requirement analysis defects, algorithm defects, document defects, performance defects, and interface defects. The technique classifies defects into classification tree nodes as indicated by defect characteristics. This technique is just considered from the development stage, and there is no code defect.

### G. Michael classification

To analyze the relationship of modules and defects, Michael proposed two classification strategies [14]: classification of defects generated by code and classification of modules in software projects. He believes that modules have a great deal to do with defect categories. For instance, interface modules are inclined to interface defects, and calculation modules are inclined to variable errors, calculation errors, etc. This classification method separates defects into eight classes: calculation type, data type, interface type, control logic type, platform type, user interface type, document type, and structure type. The strategy is accurate in classification, closely related to the module, and the logic is clear to help the developer to the defect. Numerous examinations demonstrate that the technique is a significant strategy to study the classification of software defects [15].

## IV. RESEARCH METHODOLOGY

### A. Pre-processing

There are several datasets available for software defect detection. This research used seven datasets, namely, KC1, JM1, CM1, KC2, PC1, AT, KC1 CL. All these datasets are obtained from the NASA promise dataset repository [16]. Table 1 describes the dataset in detail.

TABLE 1. CHARACTERISTICS OF THE DATASETS

| Dataset | No. of attributes | No. of instances | % of faulty instances |
|---|---|---|---|
| CM1 | 22 | 498 | 9.83 |
| JM1 | 22 | 10,885 | 19.35 |
| KC1 | 22 | 2,109 | 15.45 |
| KC2 | 22 | 522 | 20.50 |
| PC1 | 22 | 1,109 | 6.94 |
| AT | 9 | 130 | 8.46 |
| KC1 CL | 95 | 145 | 44.82 |

We performed an in-depth analysis of every dataset to understand every feature's effect on the output prediction. It was found that many features had a high correlation with others, and this led to a little overfitting. Hence, these features were not considered at the time of final training.

Principal Component Analysis (PCA) [17] is a method to reduce large datasets' dimensionality while minimizing information loss. PCA creates new uncorrelated variables such that variance is maximized. Applying this technique to the KC1 CL and JM1 results in significant improvement in performance.

### B. Models

We experimented with several machine learning models, including Logistic Regression, Random Forest [18], Naïve Bayes, Gradient Boosting Classifier, Support Vector Machine, and Artificial Neural Networks [19].

- *Logistic Regression:* It is a statistical model which uses a logistic function to model a binary dependent variable. [20]
- *Random Forest:* It is an ensemble learning method consisting of multiple decision trees and outputs the class that is the mode or mean of the trees' prediction. While building, it uses feature randomness to ensure that the individual trees are uncorrelated.
- *Naïve Bayes:* It is a probabilistic classifier based on Bayes theorem, which works on the primary assumption that features are conditionally independent [21].
- *Gradient Boosting Classifier:* It is an ensemble algorithm which is very widely used in machine learning competitions. It involves the sequential adding of models where subsequent models correct the performance of the prior model [22].
- *Support Vector Machine:* This algorithm tries to model data so that the examples of separate categories are divided by a clear gap that is as wide as possible. Apart from linear classification, they can also perform non-linear classification efficiently. [23]
- *Artificial Neural Network (ANN):* It is a computing system made by a collection of artificial neurons that are vaguely inspired by the biological brain.

The ANN was trained with a learning rate of 0.0001, which decays by a factor of 10 each time validation loss plateaus [24]. The loss function used is Binary Cross-entropy:

$$L(y, \hat{y}) = -(y * \log(\hat{y}) + (1 - \hat{y}) * \log(1 - \hat{y})) \quad (1)$$

During the training, weights of the Neural Network are adjusted to minimize the loss function. However, the magnitude and direction of weights adjustment are highly dependent on the choice of the optimizer. The most commonly used optimizer and also used in this experiment is Adaptive Moment Estimation (ADAM) [25].

## V. RESULTS

These results are obtained by 10 fold cross-validation.

Table 2. COMPARITIVE RESULTS OF ALL ALGORITHMS

| | CM1 | JM1 | KC1 | KC2 | PC1 | AT | KC1 CL |
|---|---|---|---|---|---|---|---|
| **Logistic Regression** | 85.1 | 70.6 | 80.1 | 83.7 | 79.8 | 81 | 80.5 |
| **Naïve Byes** | 82.9 | 77 | 75 | 81.7 | 81.1 | 72.7 | 73.7 |
| **Gradient Boosting Classifier** | **88** | 78.7 | **87.5** | 84 | 86 | 87 | **89.5** |
| **Support Vector Machine** | 85 | 75.4 | 83 | 86 | 85 | 88 | 78 |
| **Random Forest** | 83 | 76.9 | 85 | 79 | 89 | **91.2** | 81 |
| **ANN** | 80 | **83.4** | 83 | **88.9** | 93 | 90 | 79 |

## VI. Summary

Software defects can have a severe impact on software quality, causing problems for customers and developers. With growing complexities in software designs and technology, manual software detection becomes a challenging and time-consuming task. Thus, automatic software detection has become a hotspot for industrial research in the past couple of years. In this paper, we try to apply machine learning and deep learning to solve this problem. We use seven datasets provided by the NASA Promise dataset repository and compare the state of the art machine learning algorithms' results. This field still has much scope for improvement. We can think of some novel approaches which use complex deep learning algorithms, and also researchers should focus on more data collection.